\documentstyle[psfig,conf_iap,10pt]{article}
\def\lya{Lyman-$\alpha$}
\begin{document}
\heading{Lyman-$\alpha$ emission as a tool to study high redshift damped
systems} 
\par\medskip\noindent
\author{Olivia Puyoo and Lotfi Ben Jaffel}
\address{Institut d'Astrophysique de Paris, 98 bis Blvd Arago, 75014 Paris, 
France}

\begin{abstract}
We report a quantitative study of the escape of
Lyman-$\alpha\,$ photons from an inhomogeneous optically thick medium that 
mimics the structure of damped Lyman-$\alpha$ absorbers ({\bf{DLA}}). 
Modeling the optically thick disk with 3 components 
(massive stars and HII 
regions,
dust, and neutral hydrogen), we study
the resulting emission line profile that may arise near the extended damped
absorption profile.
\end{abstract}\\
%
%

The detection of Ly-$\alpha$ emission from DLA systems 
remains controversial \cite{hu} \cite{wo}, 
though a limited success 
has been obtained for few objects 
\cite{mo} \cite {dj} \cite{pt}. 
High-resolution profiles of emission lines contain 
information on the kinematic within the system, 
the matter spatial distribution, the dust abundance and distribution and 
the nature of the light source responsible for the emission and thus may
provide a new tool to study such objects. To evaluate this dependency, 
we derived an approximation for 
the expected width ($\Delta\lambda$\footnote{in our estimations,
we assume uniform absorption and negligible light diffusion by dust
grains.}) of 
the \lya\ line 
by using four scenarios for the Ly-$\alpha$ formation in such HI disks.

The first scenario ({\bf uniform model}) 
corresponds to a spatially connected galaxy where HI, star formation regions 
and dust are mixed. In this case the line width is classically
given by \hbox{$\Delta\lambda\simeq\sqrt\pi(\frac{N_{H20}}{2}T_4^{0.5})^{1/3}(1+{\rm
  z})\ \AA$.}

The second scenario corresponds to the {\bf lens model} \cite{mo}, in which 
the ionization of the damped system by the quasar EUV flux on its boundaries, 
results 
in Ly-$\alpha\,$ photons production by recombination processes. 
In this model, the emitting
galaxy at Ly-$\alpha\,$ appears as a bright ring due to photons escaping 
toward the observer
which should provide the galaxy spatial extension through its spectral
signature ($\Delta\lambda_{spat}$). We then obtain 
\hbox{$\Delta\lambda\simeq\,0.205\,T_4^{1/6}(1+{\rm z})+
\Delta\lambda_{spat}\,\AA$.}

The third scenario is a {\bf two phases model} 
where HI optically thick clouds 
($N(HI)\simeq\, 1.\,10^{17}\, cm^{-2}$) 
are embedded in a fully ionized region (the inter-cloud medium). 
This 
model is described by the properties of the individual HI clouds 
(opacity, individual 
albedo, temperature $T_c$, ...), of the emitting 
regions (temperature, dispersion velocity, ..), the inter-clouds velocity
dispersion $V_c$, and the dust abundance and distribution in both phases. We 
derived a good approximation of a Ly-$\alpha$ 
line emission width that may escape from such system to be 
\hbox{$\Delta \lambda \; = \; \frac{2 \lambda_o}{c} (1+{\rm z}) \left[ V_c^2 + 
\frac{2 k T_c}{m_p}
{ \left( \ln \tau_o/f_c - 0.72 \right)\over ln 2} \right]^{1/2} 
\times \left( \ln f_c/\pi + \gamma \right)^{1/2} \; \;$}
in which $f_c$ represents the line of sight covering factor, $\tau_o$ the
optical depth along the line of sight and $\gamma$ is 0.015
or 1.84 for $f_c$ larger or smaller than $10/\pi$ respectively.

In the last model, extended and emitting bubbles that may be produced by
massive stars are 
embedded in a warm HI galaxy. 
This is equivalent to an emitting layer that is embedded within the HI halo
and that has a column density  $N_{H20,s}\leq\,N_{H20}$, where $N_{H20}$ 
is the total column density along the line-of-sight. The position
within the HI region is parameterized as the column density $N_{H20,i}$ of the
intervening layer separating the emitting region from the observer. 
We derived an escaped
\lya\ line width \hbox{$\Delta\lambda\simeq\sqrt\pi(N_{H20,i}T_{4,i}^{0.5}+
\frac{N_{H20,s}}{2}T_{4,s}^{0.5})^{1/3}(1+{\rm z})\ \AA$.}
In the case the outer layer is expanding, it
is possible to show that the resulting emission line should be shrunken 
on its blue wing much like what was observed 
toward the blue compact dwarf galaxy Haro 2 \cite{le}(see Fig. 1).

\begin{figure}
\centerline{\vbox{
\psfig{figure=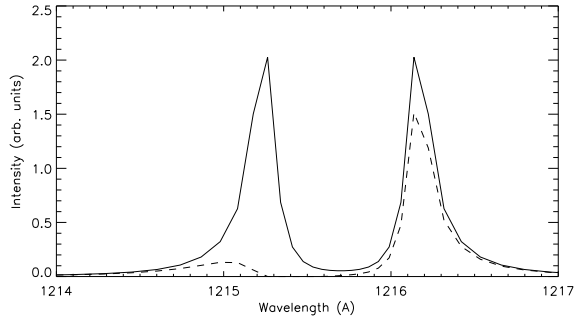,height=6.cm}
}}
\caption[]{Line profiles of the \lya\ emission assuming no wind (solid line)
 and a  $v\simeq\,75\,\, \rm km \, s^{-1}$ wind (dashed line). Note the
 residual bump on the blue wing while the core intensity is completely
 extinguished due to the absorption by the expanding shell.}
\end{figure}

%
\vfill

\begin{iapbib}{99}{
\bibitem{hu} Hunstead, R., Fletcher, A., Pettini, M., 1990, \apj 356, 23
\bibitem{wo} Wolfe, A., Lanzetta, K., Turnshek, D., 1992, \apj 385, 151
\bibitem{mo} M$\o$ller, P.,Warren, S., 1993, A\&A 270, 43
\bibitem{dj} Djorgovski, S., Pahre, M., Bechtold, J., Elston, R., 1996,
Nature 382, 234
\bibitem{pt} Petitjean, P., Pecontal, E., Valls-Gabaud, D., Charlot, S.,
1996, Nature 380, 411
\bibitem{le} Lequeux, J., Kunth, D., Mas-Hesse, J.M., Sargent, W.L., 1995,
A\&A 301, 103
}

\end{iapbib}
\end{document}